\documentstyle[sprocl,epsf]{article}

\def\Journal#1#2#3#4{{#1} {\bf #2}, #3 (#4)}

\def\PLB{{\em Phys. Lett.}  B}
\def\PRL{\em Phys. Rev. Lett.}
\def\PRD{{\em Phys. Rev.} D}

\def\be{\begin{equation}}
\def\ee{\end{equation}}
\def\bea{\begin{eqnarray}}
\def\eea{\end{eqnarray}}

\begin{document}

\title{NUCLEON SPIN CONTENT FROM ELASTIC FORM FACTOR DATA$^1$}

\author{Andrei V. Afanasev$^2$}

\address{North Carolina Central University, 
Durham, NC 27707, USA\\ and \\
Jefferson Lab, Newport News, VA 23606, USA \\E-mail: afanas@jlab.org}


\maketitle\abstracts{I use the formalism of skewed parton distributions (SPD)
to describe elastic form factors of the nucleon. The results are consistent
with approximate dipole behavior of $G_{Mp}(Q^2)$ and recent JLab data for the 
ratio of the proton form factors $G_{Ep}/G_{Mp}$ at $Q^2\leq 3.5$ GeV$^2$. Using the Angular 
Momentum Sum Rule, I obtain a numerical estimate for the valence quark contributions
to the nucleon spin at the level of 50\%, with their orbital angular momentum {\it included}. When combined with polarized DIS measurements,
this result indicated that the orbital angular momentum of quarks contributes about 25\% to the nucleon spin.}

\footnotetext{$^1$ Invited talk at the INT/JLab Workshop `Exclusive and Semiexclusive Processes at High Momentum Transfer', Newport News, VA, May 20-22, 1999}
\footnotetext{$^2$ On leave from Kharkov Institute of Physics and Technology, Kharkov, Ukraine}

\vspace{0.8cm}

\section{Introduction}

The formalism of Nonforward Parton Densities \cite{tolya} and Off-Forward
Parton Distributions \cite{amsr}, recently consolidated under a general name of Skewed Parton Distributions (SPD), provide a connection between exclusive and inclusive reactions. SPD in general describe the nonperturbatibe process of quark/parton emission and subsequent absorption by a hadron in an exclusive reaction. Since SPD assume that states of the initial and final hadrons involved in the reaction are different, they appear to be sensitive to the orbital angular momentum of quarks, and it was the original motivation of Ji to introduce this new formalism \cite{amsr}. This issue was also discussed earlier in Ref.\cite{jr}
An essential part of this formalism are helicity--flip SPD studied in Ref.\cite{hj} 

In application to specific processes, the power of SPD formalism is that 
it allows to describe universally the bag-type diagrams involving current quarks. In particular, the same SPD of a nucleon would describe nucleon elastic form factors, real and virtual Compton scattering and hard electro- and photoproduction of mesons on the nucleon. 

Here I introduce a model of helicity-flip SPD which reasonably describes available data on elastic nucleon form factors and, using Angular Momentum Sum Rule \cite{amsr}, obtain an estimate of the valence quark contribution
to the nucleon spin.

\section{A Model for SPD}

Angular Momentum Rum Rule derived by Ji \cite{amsr} relates the first moment of SPD in the forward limit to the total angular momentum of quarks and gluons. For the case of quarks it reads 
\begin{eqnarray}
J_q={1\over 2} [A_q(0)+B_q(0)],
\end{eqnarray}
where 
\begin{eqnarray}
A_q(t)=\int_{0}^{1} {\cal F}^q(x,t) x dx\\ \nonumber
B_q(t)=\int_{0}^{1} {\cal K}^q(x,t) x dx
\end{eqnarray}
are SPD using definitions of Ref.\cite{tolya}. The model for 
helicity--nonflip SPD was proposed in Ref.\cite{wacs}, assuming a Gaussian dependence of the proton light-cone wave function on the quark transverse momentum. \footnote{A similar approach was also developed in Ref.\cite{kroll}}
Complementing this picture with a model for helicity--flip SPD
${\mathcal K}^q(x,t)$, we earlier obtained a good agreement with available data on proton and neutron elastic form factors \cite{afan}. 

Key assumptions of the model are: a) helicity--flip SPD have  similar functional dependence on the transverse momentum as helicity--nonflip SPD and
b) contribution from sea quarks to the anomalous magnetic moment of the nucleon is neglected. The forward limit of helicity--flip SPD, $k_a(x)$ then enters full SPD as
\begin{eqnarray}
{\mathcal K}^a(x,t)=k_a(x) e^ {(1-x) t\over 4 x \lambda_k^2},
\end{eqnarray}
where $\lambda_k$ is an adjustable mass parameter. The distribution $k_a(x)$ cannot be observed in deep--inelastic scattering, but may be accessed in exclusive reactions, for instance, in Deeply Virtual Compton scattering \cite{tolya,amsr}.

Here I used reasonable assumptions about the function $k_a(x)$ vs. standard
parton densities (i.e., forward limit of helicity--nonflip SPD). Namely, as $x\to 1$, both pQCD and QCD Sum Rules (Ioffe current) require an extra factor of $(1-x)$ for $k_a(x)$.  For the Regge limit $x\to 0$, I explored
three possibilies a) $k_a(x)=(1-x) f_a(x)$ (Model A), b) $k_a(x)= x(1-x) f_a(x)$ and c) $k_a(x)= x^{1\over 2}(1-x) f_a(x)$  (Model C). 
The Pauli form factor of the proton is then given by the formula in terms of only valence quark distributions (contributions from sea quarks cancel):
\begin{eqnarray}
F_2(t)=\int_{0}^{1}[e_{u} k_{u}^{val}+e_{d} k_{d}^{val}] e^ {(1-x) t\over 4 x \lambda_k^2} dx.
\end{eqnarray}
\bigskip

\begin{figure}[h]
\let\picnaturalsize=N
\def\picsize{2.5in}
\def\picfilenamea{ffratio.eps}
\ifx\nopictures Y\else{\ifx\epsfloaded Y\else\input epsf \fi
\let\epsfloaded=Y
\centerline{
\ifx\picnaturalsize N\epsfxsize \picsize\fi \epsfbox{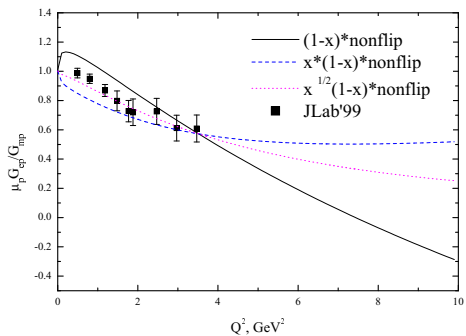}}}
\fi
\caption{Ratio of electric to magnetic proton form factors for different models of helicity--flip SPD (see text). The data are from Ref.$^8$ }
\end{figure}

\medskip

\begin{figure}[h]
\let\picnaturalsize=N
\def\picsize{2.5in}
\def\picfilenamea{gmn.eps}
\ifx\nopictures Y\else{\ifx\epsfloaded Y\else\input epsf \fi
\let\epsfloaded=Y
\centerline{
\ifx\picnaturalsize N\epsfxsize \picsize\fi \epsfbox{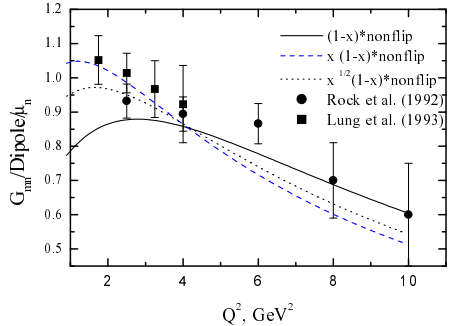}}}
\fi
\caption{Neutron magnetic form factor from SPD. Notation is the same as for Fig.1. Data are from Ref.$^{11}$.}
\end{figure}

The best fit to the recent JLab data \cite{jlab} on the ratio of proton form factors was obtained for $\lambda_k^2=0.43$ GeV$^2$ (Fig.1). For the Model A and B, this value was taken 0.64 and 0.34 $GeV^2$. These different values of $\lambda_k$ correspond to approximately the same value of about 220-250 MeV for the transverse correlation momentum defined for valence quarks as
\begin{eqnarray}
<k_{cor}^2>^a=
{\lambda_{k;a}^2\over N_a} \int_{0}^{1} x {\overline x} k^{val}_a(x) dx,
\end{eqnarray}
where $N_u=2$, $N_d$=1 are the numbers of valence $a$ quarks in the proton.
Our results for the neutron form factor (Fig.2) are consistent with data at higher $Q^2$, but the errors in measurements do not allow to discriminate between models for NFPD. We can also see a clear need for extending the measurements of the proton form factor ratio to higher $Q^2$. Fortunately, such a measurement for $Q^2=5.6$ GeV$^2$ is approved and the possibility of
measurements at 10GeV$^2$ is being studied.\cite{jlab1}

Having the model for SPD, it is straightforward to evaluate the Angular Momentum Sum Rule Eq.(1). The results for all three choices of helicity--flip SPD are presented in Table 1. Total contribution of quarks to the nucleon spin appears to be close to 50\%, with little dependence on the used model of helicity--flip SPD. This result is numerically close to the QCD Sum Rules calculation of Balitsky and Ji \cite{bj}. In our model functions $k_a(x)$ are positively defined, leading to cancellation between
up- and down-quarks for $B(0)$ in Eq.(1). In the meantime $A(0)$ is close to 0.5 -- the result known as the Momentum Sum Rule. Contributions of individual quark flavors are, however, quite sensitive to the chosen model of helicity--flip SPD. The quark total angular momentum density, 
$J_q(x)=x (f_q(x)+k_q(x))$ is plotted in Fig.3.
\begin{figure}[h]
\let\picnaturalsize=N
\def\picsize{3in}
\def\picfilenamea{spinx.eps}
\ifx\nopictures Y\else{\ifx\epsfloaded Y\else\input epsf \fi
\let\epsfloaded=Y
\centerline{
\ifx\picnaturalsize N\epsfxsize \picsize\fi \epsfbox{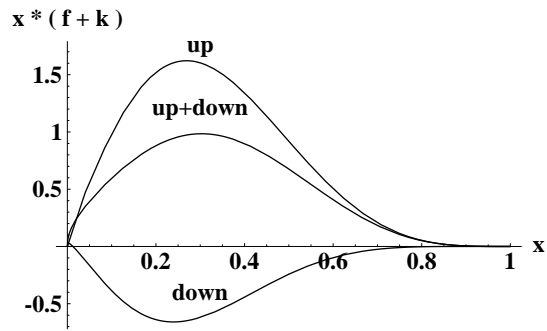}}}
\fi
\caption{Total angular momentum distribution for the quarks.}
\end{figure}

We can now combine available data from polarized deep--inelastic scattering, e.g., semi--exclusive data from SMC \cite{smc} ($\Delta u_v=0.77\pm 0.10\pm 0.08$ and $\Delta d_v=-0.52\pm 0.14\pm 0.09$) with the values from Table 1 to obtain estimates for orbital angular momentum contribution to the proton spin from individual quark flavors. The best fit to JLab data \cite{jlab} gives for the central values
\begin{eqnarray} 
\Delta L_u=(\Delta_u+L_u)^{JLab}-(\Delta u)^{SMC}=-0.04\\ \nonumber
\Delta L_d=(\Delta_d+L_d)^{JLab}-(\Delta d)^{SMC}=0.27.
\end{eqnarray}
Earlier SLAC data \cite{slac} on the proton form factor yield different numbers, $\Delta L_u=0.05$ and $\Delta L_d=0.15$.

\section{Conclusion}

I obtained new estimates of quark contributions to the nucleon spin based on formalism of Skewed Parton Distributions and using the data on nucleon elestic form factors. When combined with deep-inelastic scattering data,
it implies that about 25\% of the proton spin is due to orbital angular momentum of quarks, while other 25\% come from the quark spins. It leaves the remaining 50\% to the gluon contribution.

\begin{table}[t]
\caption{Valence Quark Contributions to the Proton Spin with
Different Models for Spin-Flip SPD (see text for notation).\label{tab:exp}}
\vspace{0.2cm}
\begin{center}
\footnotesize
\begin{tabular}{|l|c|c|l|}
\hline
{}&A&B&C\\
\hline
$\Delta_u+L_u$&0.6&0.82&0.73\\
$\Delta_d+L_d$&-0.08&-0.37&-0.25\\
$\Delta_u+L_u+\Delta_d+L_d$&0.52&0.45&0.48\\
\hline
\end{tabular}
\end{center}
\end{table}

\section*{Acknowledgments}
I would like to thank Carl Carlson and Anatoly Radyushkin for invitation to this interesting workshop. This work was partially supported by the U.S. Department of Energy under contract DE--AC05--84ER40150.


\section*{References}
\begin{thebibliography}{99}
\bibitem{tolya} A. Radyushkin, \Journal{\PLB} {380}{417}{1996}; {\it ibid.}, \Journal{\PLB} {385}{333}{1996}; \Journal{\PRD} {56}{5524}{1997}.
\bibitem{amsr} X.Ji, \Journal{\PRD}{55}{7114}{1997},  \Journal{\PRL} {78}{610}{1997}.
\bibitem{jr} P. Jain, J. Ralston, Proc. of the Workshop on
Future Directions in Particle and Nuclear Physics at multi-GeV Hadron Beam Facilities, BNL, Upton, NY,
1993.
\bibitem{hj} P. Hoodbhoy, X. Ji, \Journal{\PRD} {58}{054006}{1998}. 
\bibitem{wacs} A. Radyushkin, \Journal{\PRD} {58}{114008}{1998}.
\bibitem{afan} A. Afanasev, E-print: hep-ph/9808291. In: Proc. Workshop 
on Jefferson Lab Physics and Instrumentation with 6-12-GeV Beams and Beyond, JLab, Eds.: S.~Dytman, H.~Fenker, P.~Roos, p. 263.
\bibitem{kroll} M. Diehl et al., Phys.Lett.B460:204-212,1999; P. Kroll,
these Proceedings and E-print hep-ph/9908242.
\bibitem{jlab} M.K. Jones et al. (JLab Hall A Collab.), E-print 
nucl-ex/9910005, ({\it submitted to Phys. Rev. Lett.})
\bibitem{jlab1} {\it Measurement of $G_{Ep}/G_{Mp}$ to $Q^2$= 5.6 GeV$^2$ by the Recoil Polarization Method}, JLab experiment 99-007, Spokespersons:
C. Perdrisat, E. Brash, M. Jones, V. Punjabi; C. Perdrisat et al.,
Letter-of-Intent to JLab PAC-16.
\bibitem{bj} I. Balitsky, X. Ji, \Journal{\PRL}{79}{1225}{1997}.
\bibitem{slac} L. Andivahis et al., \Journal{\PRD}{50}{5491} {1994};
S. Rock et al.,  \Journal{\PRD} {46} {24} {1992}; A. Lung et al.,
 \Journal{\PRL} {70} {718} {1993}.
\bibitem{smc} SMC Collab., \Journal{\PLB}{420}{180}{1998}.
\end{thebibliography}
\end{document}